\documentclass[referee]{mn2e}

\usepackage{graphicx}
\usepackage{amsmath}

\title[A model for the two component gamma-ray spectra observed from the gamma-ray binaries]
{A model for the two component gamma-ray spectra observed from the gamma-ray binaries}
\author[W. Bednarek]
{W. Bednarek\\
Department of Astrophysics, University of \L \'od\'z,
ul. Pomorska 149/153, 90-236 \L \'od\'z, Poland \\
bednar@astro.phys.uni.lodz.pl}
\begin{document}

\date{Accepted . Received ; in original form }

\pagerange{\pageref{firstpage}--\pageref{lastpage}} \pubyear{2007}

\maketitle

\label{firstpage}

\begin{abstract}
Observations of two well known binary systems (LS 5039 and LS I +61 303) with the satellite and Cherenkov telescopes revealed the broad band $\gamma$-ray spectra which seems to show two components, the first at GeV energies, showing exponential cut-off at a few GeV, and the second at TeV energies which does not fit well to the extrapolation of spectrum from the GeV energy range. We propose that such two component spectrum is produced by two populations of electrons which appear within the binary system as a result of acceleration on a double shock structure separated by a contact discontinuity. Such structure is created within the binary system as a result of the interaction of the pulsar and massive star winds. The shocks from the side of the pulsar and the massive star have different proprieties which allow acceleration of electrons to different maximum energies. These two populations of electrons produce 
two component $\gamma$-ray spectra caused by the Inverse Compton (IC) scattering of stellar radiation. 
The example calculations, performed in terms of the anisotropic IC $e^\pm$ pair cascade model, for the location of the pulsar at the periastron and apastron passages confirm the high energy emission features observed from LS I +61 303. 
\end{abstract}

\section{Introduction}

The possibility of high energy $\gamma$-ray production within the massive binary systems containing energetic pulsars has been suspected since early 80'-ties (see e.g. Maraschi \& Treves~1981, Arons \& Tavani~1993, Kirk et al.~1999, Kawachi et al.~2004, Sierpowska \& Bednarek 2005).  
However, only recently three gamma-ray binaries have been well documented as $\gamma$-ray sources at first in TeV energies (LS 5039 - Aharonian et al. 2005a; LS I +61 303 - Albert et al.~2006; SS2883/PSR 1259-63 - Aharonian et al. 2005b) and also later in GeV energies (LS 5039 - Abdo et al. 2009a; LS I +61 303 - Abdo et al. 2009b, Tavani et al.~2009; SS2883/PSR 1259-63 - Abdo et al. 2010a, Tavani et al.~2010). $\gamma$-ray observations suggest that there are two components in the broad band $\gamma$-ray spectra from these sources. The $\gamma$-ray spectra of LS 5039 and LS I +61 303 show the cut-offs at a few to several GeV. These cut-offs do not fit nicely to the higher level of the TeV emission observed from these two sources.  
With the discovery of the binary system SS2883/PSR 1259-63 and evidences of characteristic radio tail within the binary system LS I +61 303 (Dhawan et al.~2006), the general scenario involving energetic pulsar has been considered in detail for the $\gamma$-ray binary systems (e.g. Dubus 2006, Romero et al.~2007, Sierpowska-Bartosik \& Torres 2007, 2009, Zdziarski et al.~2010).

The level of GeV $\gamma$-ray emission from these binary systems might be consistent with the contribution from the inner pulsar magnetosphere assuming that the pulsar has the parameters comparable to those observed in the case of the Vela type pulsars (see e.g. Abdo et al. 2010). However, such emission should be steady, without any evidences of strong variability with the phase of the binary system (clearly observed in the case of LS 5039 and LS I +61 303). Moreover, $\gamma$-ray emission should be also detected far away from the periastron passage which is not the case of the binary system PSR 1259-63/SS 2883 (see Tam et al.~2011,
Abdo et al.~2011). Therefore, this interpretion seems to be inconsistent with the basic observational facts. Here we consider another scenario which has been recently proposed as a possible explanation of the $\gamma$-ray emission from the binary system of two massive stars Eta Carinae (Farnier et al.~2010, Bednarek \& Pabich~2011). We assume that two populations of relativistic electrons appear within the binary systems containing pulsars as a result of their acceleration on a double shock structure created as a result of collisions of the pulsar wind and the stellar wind (see also Arons \& Tavani~1993, Bednarek \& Pabich~2011). These shocks are able to accelerate particles to different energies due to different plasma proprieties. Note that possibility of acceleration of particles on both shocks has been already considered as a possible explanation of the non-thermal X-ray emission from the binary systems containing a millisecond pulsars (e.g. PSR 1957+20, (Arons \& Tavani~1993) and a classical pulsar (SS2883/PSR1259-63, Tavani \& Arons 1997).

\section{Colliding winds within binary system}

We consider a classical scenario in which an energetic pulsar (of the Vela type) orbits the massive companion star of the O or B type. Both stars create strong winds which collide within the binary system. The shock geometry is determined by the dimensionless parameter $\eta$ (e.g. Ball \& Dodd, 2000) defined as follows:
$\eta =  L_{\rm rot} /(\dot{M} V_{\rm w}c)$,
where $V_{\rm w} = 10^8V_8$ cm s$^{-1}$ is the stellar wind velocity, $\dot{M}$ is the mass loss rate, $c$ is the velocity of light, and $L_{\rm rot}$ is the rotational energy loss rate of the pulsar estimated from 
\begin{eqnarray}
L_{\rm rot}\approx 6\times 10^{35} B_{12}^2 P_{ms}^{-4}~{\rm erg}~{\rm s}^{-1},
\label{eq2} 
\end{eqnarray}
\noindent
where $P = 100P_{\rm 100}$ ms is the rotational period of the pulsar and $B = 10^{12}B_{\rm 12}$ G is its surface magnetic field strength. 
The closest radial distance of the shock from the pulsar (measured in the plane of the
binary system) is then equal to,
$R_{\rm pul}^{\rm sh} =  D \ \sqrt{\eta}/(1+\sqrt{\eta})$  and
$R_\star^{\rm sh} =  D/(1+\sqrt{\eta})$ from the massive star, where $D = (1 \pm e)a$ ($a$ semimajor axis and $e$ eccentricity) is the separation of the stars at the apastron and periastron passages, respectively. 

As a result of the collision between the winds, a double shock structure appears separated by a contact discontinuity (see Fig.~1). The conditions at the shocks created at the sides from the pulsar and the massive star differ significantly. 
The shock from the pulsar is relativistic and contains $e^\pm$ plasma from the inner pulsar magnetosphere which arrives to the shock with large Lorentz factor (of the order of $10^{4-6}$, Kennel \& Coroniti~1984). On the other hand, the shock from the side of the massive star is non-relativistic and contains electrons with low initial energies. It is expected that the acceleration of particles on both shocks occurs differently due to differences in the magnetic field strengths and the acceleration efficiencies defined by the so called acceleration parameter $\xi$. The value of this parameter is often related to the velocity of the plasma through the shock by $\xi\sim (v_{\rm w}/c)^2$. Below we estimate the maximum energies of electrons accelerated in both shocks (see also recent paper by Bednarek \& Pabich~2011).

\begin{figure}
\vskip 5.6truecm
\includegraphics{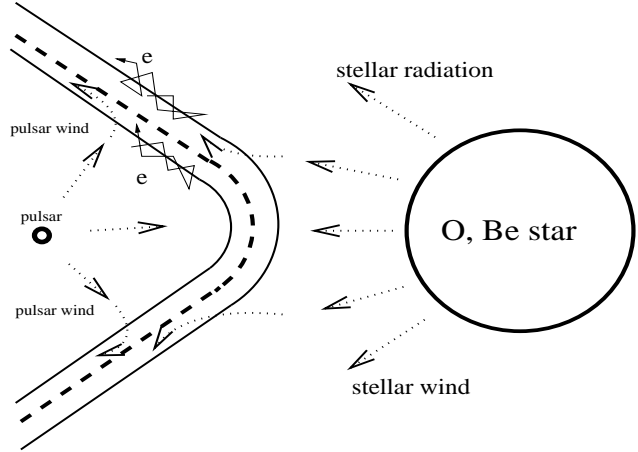}
\caption{Schematic representation of the massive binary system containing an energetic pulsar and a massive star. As a result of the stellar wind and pulsar wind collision a double shock structure appears separated by a contact discontinuity. The physical parameters on both sides of the contact discontinuity differ significantly allowing acceleration of electrons to different maximum energies. Accelerated electrons comptonize the stellar radiation from the massive star. If electron energies are large enough they can initiate IC $e^\pm$ pair cascade in the radiation field of the massive star.}
\label{fig1}
\end{figure}

The acceleration rate of particles at the shock can be parametrised by
${\dot P}_{\rm acc} = \xi cE/R_{\rm L}\approx 10^4\xi B~{\rm GeV~s^{-1}}$
where $E$ is the energy of particles (in GeV), and $R_{\rm L}$ is the Larmor radius of particles in the magnetic field $B$ (in Gauss). We estimate the characteristic acceleration time scale of particles from
\begin{eqnarray}
\tau_{\rm acc} = E/{\dot P}_{\rm acc}\approx 10^{-4}E/(\xi B)~~~{\rm s}.
\label{eq2}
\end{eqnarray}
\noindent
The acceleration process is limited by the energy losses of electrons through different radiation processes. We estimate the energy loss time scales for electrons and their maximum allowed energies at the shocks from the sides of the pulsar and the companion star. The magnetic field at the shock region from the pulsar side is estimated by assuming that it has a dipole structure in the inner pulsar magnetosphere (below the light cylinder radius, $B\propto D^{-3}$) and has a toroidal structure in the pulsar wind region (above the light cylinder, $B\propto D^{-1}$). Then, the magnetic field strength at the shock region is
\begin{eqnarray}
B^{\rm sh}_{\rm pul}\approx 0.44\sigma_{-2}^{1/2}B_{12}/(P_{100}^{2}D_{12})~~~{\rm G},
\label{eq4}
\end{eqnarray}
\noindent
where $\sigma = 10^{-2}\sigma_{-2}$ is the pulsar wind magnetization parameter (see e.g. Kennel \& Coroniti~1984), and the distance of the shock from the pulsar is $D = 10^{12}D_{12}$ cm.
The magnetization parameter is estimated in the case of the Crab Nebula on $\sim 3\times 10^{-3}$. However, in the case of the shocks which are either closer to the pulsar, or the pulsar of the Vela type, the value of $\sigma$ can be larger. In our calculations we assume $\sigma$ of the order of $\sim 10^{-2}$.

On the other hand, the magnetic field strength at the shock from the massive star has dominating radial structure (we neglect inner dipolar structure since it extends only close to the stellar surface). Then, the magnetic field from the side of the massive star can be estimated from
\begin{eqnarray}
B^{\rm sh}_\star = 100B_{100}/D_{\rm sh}^2~~~{\rm G},
\label{eq5}
\end{eqnarray} 
where the surface magnetic field of the massive star is $B_\star = 100B_{100}$ G, and the distance of the shock from the center of the star is expressed in stellar radius ($R_\star$), i.e. $D = D_{\rm sh}\cdot R_\star$.
Note that for the expected values of the magnetic fields on the surface of the pulsar and the star the value of the magnetic field strength on both sides of the double shock structure and the efficiency of acceleration of particles, $\xi$, can differ significantly. 
In the case of the shock in the pulsar wind the plasma still moves relativistically after the shock with the velocity of the order of $\sim 0.3c$. So then, for this shock 
we estimate $\xi_{\rm pul}\sim 0.1$. On then other hand, the plasma in the wind of the massive star moves much slower with a characteristic velocity of the order of $v_\star\sim 10^3v_3$ km s$^{-1}$. In such a case the acceleration parameter is of the order of $\xi_\star\sim (v_\star/c)^2\sim 10^{-5}v_3^2$. These different conditions at the shocks from the pulsar and the stellar sides should result in very different acceleration time scales of particles. 

The cooling time scales of electrons on the synchrotron and the IC (in the Thomson regime) processes can be estimated from,
$\tau_{\rm syn/IC} = E_{\rm e}/{\dot P}_{\rm syn/IC}$, where $E_{\rm e}$ is the electron energy and ${\dot P}_{\rm syn/IC}$ are the energy loss rates of electrons on both processes. These time scales are estimated as,
\begin{eqnarray}
\tau_{\rm syn} = {{E_{\rm e}m_{\rm e}^2}\over{4/3c\sigma_{\rm T}U_{\rm B}E_{\rm e}^2}}\approx {{3.7\times 10^5}\over{B^2E_{e}}}~~~{\rm s},
\label{eq6}
\end{eqnarray}
\noindent
and
\begin{eqnarray}
\tau_{\rm IC/T} = {{E_{\rm e}m_{\rm e}^2}\over{4/3c\sigma_{\rm T}U_{\rm rad}E_{\rm e}^2}}\approx {{170D_{\rm sh}}\over{E_{\rm e}T_4^2}}~~~{\rm s},
\label{eq7}
\end{eqnarray}
\noindent
where $\sigma_{\rm T}$ is the Thomson (T) cross section, $U_{\rm B}$ and $U_{\rm rad}$ are the energy densities of the magnetic and radiation fields, $m_{\rm e}$ is the electron mass, and $T = 10^4T_4$K is the surface temperature of the massive star. The cooling time scale in the Klein-Nishina (KN) regime can be roughly estimated by introducing the value of energy of electrons, into the above formula for the IC losses, corresponding to the transition between the T and the KN regimes,
$E_{\rm e}^{\rm T/KN} = m_{\rm e}^2/(3k_{\rm B}T)\approx 97/T_4$ GeV. 
Then, the cooling time scale of electrons in the KN regime in the radiation field of both stars is approximately given by
\begin{eqnarray}
\tau_{\rm IC}^{\rm KN} = {{3E_{\rm e}m_{\rm e}^2}\over{4c\sigma_{\rm T}U_{\rm rad}(E_{\rm e}^{\rm T/KN})^2}}\approx 0.27E_{\rm e}R_{\rm sh}^2/T_4^4~{\rm s}.
\label{eq8}
\end{eqnarray}
\noindent
The maximum energies of accelerated electrons should be determined by balancing the energy gains from the acceleration process with the energy losses on the synchrotron and IC processes. The synchrotron losses dominates over the IC losses even in the T regime for the magnetic field strength at the shock, 
$B^{\rm sh} > B^{\rm sh}_{\rm T} = 40T_4^2/D_{\rm sh}~{\rm G}$.
When the acceleration energy gains overcome the IC energy loss rate in the KN regime, then the saturation of acceleration process is due to the synchrotron energy losses. This occurs for the magnetic field strength fulfilling the condition,
$B^{\rm sh}_{\rm syn} > 4\times 10^{-4}T_4^4/(\xi D_{\rm sh}^2)~~~{\rm G}$. 
So then, depending on the parameters of the massive star, its distance from the shock, and the acceleration efficiency, we can distinguish following situations,

\begin{enumerate}

\item $B_{\rm sh} > B^{\rm sh}_{\rm T}$ and $B_{\rm sh} > B^{\rm sh}_{\rm syn}$: acceleration of electrons is always saturated by synchrotron energy losses at energies above $E_{\rm e}^{T/KN}$. 
This likely happen at the shock from the pulsar side. In this case electrons lose efficiently energy on the IC and synchrotron processes. The 
maximum energies of electrons are obtained by balancing acceleration time scale with the cooling time scale on synchrotron process. They are equal to 
\begin{eqnarray}E_{\rm e}^{\rm max}\approx 63(\xi/B)^{1/2}\approx 90P_{\rm 100}\mu_{-2}^{-1/4}(\xi D_{12}/B_{12})^{1/2}~~~{\rm TeV}.
\label{eq11}
\end{eqnarray}
\noindent
Note that the maximum energies of electrons accelerated at the shock from the pulsar side increase with the distance of the shock from the massive star as $\propto D^{1/2}$. 

\item $B^{\rm sh}_{\rm T} > B_{\rm sh} < B^{\rm sh}_{\rm syn}$: acceleration of electrons is saturated by IC energy losses in the T regime. 
This likely happens for electrons accelerated at the shock from the side of the massive star. Then, their maximum energies are equal to 
\begin{eqnarray}
E_{\rm e}^{\rm max}\approx 1.3(\xi B_{\rm sh})^{1/2}(D_{\rm sh}/T_4^2)\approx 13\xi^{1/2}B_{\rm 100}/T_4^2~~~{\rm TeV}.
\label{eq9}
\end{eqnarray}
\noindent
Note that in this case the maximum energies of accelerated electrons does not depend directly on the distance of the shock from the massive star. They are constant for different parts of extended shock structure or the shocks at different phases of the binary system.

\item $B_{\rm sh} > B^{\rm sh}_{\rm T}$: acceleration of electrons is saturated by the synchrotron energy losses (as in the case (i)) at energies below $E_{\rm e}^{T/KN}$. However, these electrons lose most of their energy on synchrotron radiation producing low energy photons. 

\end{enumerate}

In summary, electrons should be accelerated to TeV energies at the
shock from the pulsar side and to several GeV energies at the shock from the massive star side.
We assume that in both cases electrons obtain the power law spectra up to the maximum energies estimated from the above formulae. However, their acceleration rate may be also different on both sides of the contact discontinuity. Note that leptons injected into the acceleration mechanism from the pulsar side already have large Lorentz factors due to the confinement in the relativistic pulsar wind.

The maximum power in relativistic electrons is limited by the power of the pulsar and stellar winds. In the case of the pulsars in these binaries we applied the parameters 
($B_{12} = 3$, $P_{100} = 0.5$) which result in the total energy of the pulsar wind $\sim$$8\times 10^{37}$ erg s$^{-1}$, i.e. between those observed from  the Vela and Crab pulsars. On the other hand, the power of the stellar wind can be estimated from $L_{\rm w} = 0.5{\dot M}v_{\rm w}^2\approx 5\times 10^{35}$ erg s$^{-1}$. This seems to be marginally enough to supply the $\gamma$-ray luminosity of these binary systems specially when we take into account that the $\gamma$-ray emission from these systems is expected to be highly aspherical as suggested by the observed $\gamma$-ray light curves (Abdo et al.~2009a,b) and also the cascade calculations (e.g. Bednarek~2006).

%%%%%%%%%%%%%%%%%%%%%%%%%%%%%
%\begin{quote}
\begin{table*}
  \caption{Basic parameters of the stars in $\gamma$-ray binary systems and shock localizations  (distances in $10^{12}$ cm)}
  \begin{tabular}{lllllllllllll}
\hline 
\hline 
\\
Name   &  $B_{\rm pul}$ (G) &  $P_{\rm pul}$ (s) &  ${\dot M}$ $({{M_\odot}\over{yr}})$ & $v_{\rm w}$ (${{km}\over{s}}$) & a  & e & $\eta$  & $R_{\star , p}^{\rm sh}/R_{\star , a}^{\rm sh}$  & $T_\star$ (K) & $R_\star$ \\
       &           &       &    & & & &  & &  \\
\hline
\\
LS 5039        &  $3\times 10^{12}$   &   $0.05$  &  $2.6\times 10^{-7}$ & $2.4\times 10^3$   & $2.3$ & 0.35  & ~0.7  & $0.8$/$1.6$  & $3.9\times 10^4$ &  $0.65$\\
\\
LS I +61 303   &  $3\times 10^{12}$         &  0.05  &  $3\times 10^{-7}$   &  $2\times 10^3$  & $6.3$ & 0.55  & ~0.7 & $1.9$/$5.4$ &  $3\times 10^4$   & $0.47$ \\
\\
%PSR 1259-63    &  $3\times 10^{11}$ &  0.048   &     &  & $75$ &  0.87 &  
%&    & $2.7\times 10^4$  &  $0.42$ \\
%(poloidal)    &    &   &  $2\times 10^{-10}$   &  $1.5\times 10^3$  &  &  &
% ~17 & $1.9$  &  \\
%(equatorial)   &   &  &  $2\times 10^{-7}$   &  $100$ &  &  & ~0.017 &  $8.6$  &  \\
% \\
\hline
\hline 
\end{tabular}
  \label{tab1}
\end{table*}
%\end{quote}
%%%%%%%%%%%%%%%%%%%%%%%%%%%%%%%%%

%
%
\section{Application to gamma-ray binaries}

The basic parameters of the massive stars in the $\gamma$-ray binary systems LS5039 and LS I +61 303 does not differ drastically (see collection in Table 1 based mainly on Grundstrom et al.~(2007) and Casares et al.~(2005)). These parameters allow us to estimate the closest radial distance of the shock structure to the center of the massive star at the periastron and apastron passages, i.e. $R^{\rm sh}_{\star, p}$ and $R^{\rm sh}_{\star, a}$. The proprieties of the electron spectra in these two binaries and conditions for their propagation are also quite similar. We assume that the pulsars in these two binaries are of the Vela type and have similar parameters. 
Based on the formulae derived above, we estimate the basic parameters characterizing the considered scenario such as: the maximum energies of electrons at the periastron and apastron passages, the magnetic field strength at the shocks, and the distance of the shocks from the massive stars (see Tables 1 and 2). Their knowledge allow us to calculate the $\gamma$-ray spectra expected from these binary systems.
Note that electrons accelerated at the shocks from the side of the massive star reach maximum energies of the order of $\sim$10 GeV. These energies are independent on the shock distance from the star (see Eq.~9). Electrons accelerated at the shock from the side of the pulsar obtain the maximum energies of the order of $\sim$10 TeV. 
These energies increase proportionally to the square distance of the shock from the 
pulsar. Therefore, they are larger at the apastron passage.

%%%%%%%%%%%%%%%%%%%%%%%%%%%%%
%\begin{quote}
\begin{table}
  \caption{Acceleration efficiencies at the shocks from the stars and the maximum energies of accelerated electrons in both shocks}
  \begin{tabular}{llllllllllllll}
\hline 
\hline 
\\
Name   &  $\xi_\star$  &  $E_\star$   & $E_{\rm pul}^{\rm peri}$   &  $E_{\rm pul}^{\rm apo}$  \\
       &           &       &    &   \\
\hline
\\
LS 5039        & $6\times 10^{-5}$  & $6.8$ GeV &   6.8 TeV & 10.5 TeV \\
\\
LS I +61 303  & $4\times 10^{-5}$   & $11$ GeV & 10 TeV & 17 TeV\\
\\
%PSR1259-63     &   &   &     &     \\
%(poloidal)   &  $2.5\times 10^{-5}$  & 8 GeV  & 24 TeV &  ~~~x \\
%(equatorial)    &  $10^{-7}$   &  0.3 GeV   & 3.6 TeV  &  ~~~x \\
%\\
\hline
\hline 
\end{tabular}
  \label{tab2}
\end{table}
%\end{quote}
%%%%%%%%%%%%%%%%%%%%%%%%%%%%%%%%%

For the illustration, we have performed the calculations of the $\gamma$-ray spectra produced at the periastron and apastron passages of the pulsar applying the parameters of the binary system LS I +61 303. In our calculations, electrons are injected at both sides of the shock structure. Leptons accelerated to TeV energies at the pulsar shock comptonize the soft radiation from the massive star initiating the IC $e^\pm$ pair cascade. We follow this cascade process by applying the cascade model described in a more detail in Bednarek \& Giovannelli~(2007). This code also include the synchrotron energy losses of primary electrons. The aim of these example calculations is to show that the features of the $\gamma$-ray spectra expected in such a model are generally consistent with the $\gamma$-ray observations of the binary system
LS I +61 303. We realize that the details of the spectrum may depend on many not well known parameters, e.g. the exact shape of the double shock structure, dependence of the injection efficiency and the spectrum of electrons on the location in different parts of the shock and at different phases, possible existence of the aspherical wind and its geometry (expected in the case of LS I +61 303), the unknown parameters of the pulsar and the acceleration process (e.g. dependence of acceleration efficiency on 
the location at the shock). These parameters are impossible to fix precisely enough at the present stage of our knowledge. Therefore, any detailed fitting of the 
$\gamma$-ray spectra of these sources are premature.  
In our calculations we have fixed the surface magnetic field strength of the massive star on 100 G which seems to be typical for the the O and B type stars. The Lorenz factor of the pulsar wind at the distance of the shock has been fixed on $\gamma_{\rm w} = 2\times 10^5$. It is assumed that leptons injected into the shock acceleration process have such minimum energies, i.e. $E_{\rm min} = m_{\rm e}\gamma_{\rm w} = 100$ GeV. The maximum energies of leptons accelerated at the shock from the pulsar side can be in principle even significantly larger (than estimated in Tab.~2), in the case of faster rotating pulsar with weaker surface magnetic field or if the significant reconnection of the magnetic field occurs in the pulsar wind resulting in the lower value of the magnetization parameter (see e.g. Sierpowska \& Bednarek 2005).

The results of these example calculations are shown in Fig.~2 for the injection region located at the closest part of the shock to the massive star, i.e. $R_{\star p}^{\rm sh}$ and $R_{\star, a}^{\rm sh}$, and assuming this same spectral index for both populations of electrons equal to -2. 
We do not have at present enough knowledge to conclude what is the injection rate of relativistic electrons at different parts of the shock structure and at different phases of the system. Therefore, it is rather difficult to predict the exact shape of the $\gamma$-ray spectrum producted by electrons accelerated at the extended shock region. The calculations are performed for the inclination angle of the binary system
equal to $i = 30^{\rm o}$ and the longitude of the periastron $\omega_{\rm p} = 57^{\rm o}$ (Grundstrom et al.~2007).

As expected, two populations of electrons produce two distinct components in the $\gamma$-ray spectrum. The GeV flux can change by a factor of a few even in the case of this same injection rate of electrons at the shock from the side of the massive star. On the other hand, the TeV flux does not show similar variability between periastron and apastron passages. This can be understood in terms of the considered 
IC $e^\pm$ pair cascade model which produce more anisotropic emission at GeV energies (more concentrated on the directions close to the massive star) than at TeV energies (see e.g. Bednarek~2000).
 
Note that the $\gamma$-ray spectra produced by electrons accelerated at the shock from the side of the massive star fits reasonably well to the  
synchrotron spectrum produced by primary leptons accelerated at the shock from the side of the pulsar.
For the considered parameters, the magnetic field strength at the shocks from the side of the pulsar and the massive star are comparable in the case of LS I +61 303 (of the order of $\sim 5$ G at the periastron and $\sim 1$ G at the apastron). Therefore, the synchrotron emission from the secondary cascade $e^\pm$ pairs should not completely dominate the synchrotron emission from the primary electrons. The relative importance of the synchrotron emission from secondary cascade $e^\pm$ pairs should be even lower for electrons accelerated at the parts of the shock which are farther from the massive star due to lower average magnetic field strength in the massive star wind.

\begin{figure}
\vskip 4.9truecm
\includegraphics{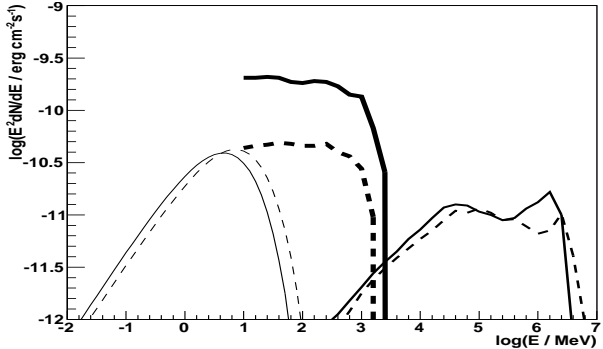}
\caption{The example $\gamma$-ray spectra calculated in terms of the two component model in which two population of electrons appear as a result of acceleration on the shock from: the side of the pulsar (middle thick curves) and the side of the massive star (thick curves). The synchrotron spectra from primary electrons accelerated at the pulsar shock are marked by the thin curves. It is assumed that electrons are injected at the parts of the shocks which are the closest to the massive star. The spectra of electrons are of the power law type with the spectral index equal to $-2$ and the cut-off as estimated in Table.~2. The electron spectra at the periastron and the apastron passages are normalized to this same injection rate. Produced by them $\gamma$-ray spectra are shown by the solid and dashed curves, respectively. The calculated $\gamma$-ray spectra have been re-normalized to the flux observed by the Fermi (Abdo et al.~2009a) and the MAGIC telescopes (Albert et al.~2009) (.}
\label{fig2}
\end{figure}

The spectra presented in Fig.~2 are obtained by assuming that relativistic electrons can cool locally on the IC and synchrotron processes to energies $\sim$100 MeV, i.e. close to the acceleration place in the shock. 
In fact, the cooling time scale of electrons in the radiation field of the star is much shorter than the dynamical time scale of the pulsar on its orbit around the star, $\tau_{\rm dyn}\approx D/v_{\rm orb}$, where $v_{\rm orb}$ is the velocity of the pulsar on its orbit. Also the advection time scale of electrons along the shock, $\tau_{\rm adv}\approx 3R_\star^{\rm sh}/v_{\rm w}$, is much longer than the IC cooling time scale.

As we already noted, important complication to the considered scenario can be introduced by the aspherical wind of the massive star, as expected in the case of LS I +61 303 or SS2883/PSR1259-63 (see Sierpowska-Bartosik \& Bednarek~2008).
The aspherical winds can significantly change the localization of the shock structure
for different phases of the pulsar
resulting in different parameters at the shock and consequently different maximum energies of electrons and possibly also their injection rates. This effects can 
influence the observed $\gamma$-ray light curves from these binary systems at GeV and TeV energy ranges.

Note that also ions can be accelerated at  the shock from the side of the massive star. However they are not able to lose efficicently energy in hadronic collisions with the matter of the stellar wind during their advection time along the shock surface for the parameters of LS +61 303.

\section{Conclusion}

We propose a simple model for the explanation of the two-component $\gamma$-ray spectra observed from $\gamma$-ray binaries (LS 5039 and LS I +61 303), presumably containing energetic pulsars, in which electrons are accelerated to different maximum energies at the shocks in the pulsar and stellar winds. We show that for the maximum energies of electrons, accelerated at the shock from the side of the massive
star, should not depend on the distance from the star (and so on the phase of the binary system). As a consequence, the cut-off in the GeV part of the spectra should 
be independent on the phase of the binary. On the other hand, the maximum energies 
of leptons accelerated at the shock from the pulsar side depend on the distance as 
$\propto D^{1/2}$. Thus, the TeV $\gamma$-ray spectra should extend to larger energies at the apastron passage. In the case of injection rate of electrons independent of the phase, the model predicts strong variability at GeV energies at much lower variability at TeV energies for the parameters of the binary system LS I +61 303. 
We comment that the detailed modelling of the $\gamma$-ray light curves and spectra from these $\gamma$-ray binary systems is at present very difficult since it depends on many parameters which are not known with enough precision.

\section*{Acknowledgments}
This work is supported by the Polish MNiSzW grant 745/N-HESS-MAGIC/2010/0.

%%%%%%%%%%%%%%%%%%%%%%%%%%%%%%%%%%%%%%%%%%%%

\label{lastpage}
\end{document}